\documentclass[aps,pra,preprint,groupedaddress,showpacs]{revtex4}
\usepackage{graphicx}
\usepackage{psfig}
\usepackage{epsf}
\usepackage{epsfig}
\usepackage{amsmath}
\begin{document}

\title{Kinetic modelling and molecular dynamics simulation of ultracold neutral
plasmas including ionic correlations}

\author{T.\ Pohl}
\author{T.\ Pattard}
\author{J.M.\ Rost}

\affiliation{Max Planck Institute for the Physics of Complex Systems,
N{\"o}thnitzer Str.\ 38, D-01187 Dresden, Germany}

\date{\today}

\begin{abstract}
A kinetic approach for the evolution of ultracold neutral plasmas including
interionic correlations and the treatment of ionization/excitation and
recombination/deexcitation by rate equations is described in detail.
To assess the reliability of the approximations inherent in the kinetic model,
we have developed a hybrid molecular dynamics method. Comparison of the results
reveals that the kinetic model describes the atomic and ionic observables of
the ultracold plasma  surprisingly well, confirming
our earlier findings concerning the role of ion-ion correlations  [Phys.\ Rev.\
A {\bf 68}, 010703]. In addition, the molecular dynamics approach allows one to
study the relaxation of the ionic plasma component towards thermodynamical
equilibrium.
\end{abstract}


\maketitle

\section{Introduction} \label{intro}
Recent experiments have produced ultracold neutral plasmas 
from a small cloud of laser-cooled atoms confined in a magneto-optical trap
\cite{Kil99,Kul00,Kil01,Rob00,Eyl00,Gou03}. In one type of experiments
\cite{Kil99,Kul00,Kil01}, a plasma was produced by photoionizing
laser-cooled Xe atoms with an initial ion temperature of about $10 \,
\mu {\rm K}$.
By tuning the frequency of the ionizing laser, the initial electron energy
$E_{\rm e}$ could be varied corresponding to a temperature range $1 \, {\rm K}
< E_{\rm e} / k_{\rm B} < 1000 \, {\rm K}$, and the subsequent expansion of the
plasma into the surrounding vacuum was studied systematically. In a
complementary type of experiment \cite{Rob00,Eyl00,Gou03}, ultracold Rb and Cs
atoms were laser-excited into high Rydberg states rather than directly ionized.
In these experiments, also the spontaneous evolution of the Rydberg gas into a
plasma has been observed.
The time evolution of several quantities characterizing the
state of the plasma, such as the plasma density \cite{Kil99,Kul00}, the degree
of ionization \cite{Rob00,Eyl00,Gou03} or the energy-resolved atomic level
population \cite{Kil01} have been measured using various plasma diagnostic
methods.

These experiments, which have paved the way towards an unexplored regime of
ionized gases, give rise to new phenomena in atomic physics as well as in
plasma physics. Hence, a number of different theoretical approaches have
been formulated to cover different aspects of these experiments 
\cite{Kuz02,Kuz02b,Maz02,Rob02,Rob03,Tka01,PPR03}. 
 
An important issue is the question whether the plasma produced would be
strongly coupled or not.
The correlation strength is determined by the Coulomb coupling
parameter $\Gamma=e^{2}/(a k_{{\mathrm{B}}} T)$ with the Wigner-Seitz radius
$a$ \cite{Dub99}. A plasma is called ``strongly coupled'' if $\Gamma \gg 1$,
i.e.\ if the Coulomb interaction between the plasma particles largely exceeds
the thermal kinetic energy. In this case, interesting ordering effects such as
Coulomb crystallization can be observed \cite{Ich82,PPR04}.
For the initial conditions of the NIST experiments
\cite{Kil99,Kul00,Kil01}, however, the development of equilibrium
electron-electron
correlations leads to a rapid heating of the electron gas, which prevents the
electron Coulomb coupling parameter $\Gamma_{\rm e}$ from exceeding unity
\cite{Kuz02}. The same has been argued for an ion plasma in \cite{Mur01}.
Since the electron dynamics proceeds on a much smaller timescale than the ion
motion, in \cite{Kuz02,Kuz02b} the electron heating could be studied for the
early stage of the plasma evolution only, where the ionic component does not
show dynamical effects. On the other hand, ion heating has only been studied in
the framework of a model system, consisting of a homogeneous gas of
Debye-screened ions \cite{Mur01}, such that the influence of the subsequent
expansion could not be explored. 

The first
quantitative comparison with experimental observations has been given in
\cite{Rob02}, with the plasma dynamics  modeled within a
hydrodynamical approach and ionization, excitation and recombination
treated by a separate set of rate equations.
Since this model does only account for
the mean-field potential created by the charges, it cannot describe effects of
particle correlations. However, it has also been shown there that the
electronic Coulomb coupling parameter $\Gamma_{\rm e}$
does not exceed a value of $\approx 0.2$ during the plasma expansion
due to heating  by three-body recombination. Thus,
the influence of electron-electron correlations on the dynamics of the plasma
could be neglected. On the other hand, three-body recombination does not
influence the ionic temperature, so that the ions can heat up only through
correlation heating (and energy exchange with the electrons, which, however,
is very slow). Since the ionic temperature was set to zero in \cite{Rob02},
the role of ion-ion correlations could not be explored. In a
preliminary study \cite{PPR03}, we showed that they indeed change the
evolution of the system quantitatively, though not qualitatively. In the
following, we will give a detailed account of the kinetic model used in
\cite{PPR03} and of all relevant
ingredients. We will also develop a hybrid molecular dynamics (H-MD) approach
which treats the electronic plasma component in an adiabatic approximation while
the ions are fully accounted for. Such an approach permits the description of
situations where the ions are strongly coupled \cite{PPR04,PPR04b}, which is
clearly beyond the capabilities of the simple kinetic model. Nevertheless, for
the typical situations realized in the experiments \cite{Kil99,Kul00,Kil01},
comparison of the two theoretical approaches yields very good agreement,
corroborating our findings reported earlier \cite{PPR03} and establishing
firmly that one can 
capture the relevant physics with the relatively simple kinetic approach.
\section{Theoretical Approach}
Our kinetic approach is similar to the one of \cite{Rob02}.
The main difference is the inclusion of ion-ion correlations (IC) which will be
described in detail below. Briefly, a set of kinetic equations is formulated for
the evolution of the plasma (subsection \ref{za}), while ionization/excitation
and recombination/deexcitation are taken into account on the basis of rate
equations (subsection \ref{zb}). In order to test the applicability and
accuracy of this model, we have developed a less approximative and more
flexible but computationally much more demanding approach. It uses 
molecular dynamics for the ionic motion while the electron component is
treated as a fluid assuming a quasi-steady state (subsection \ref{zc}). 

\subsection{Kinetic description}
\label{za}
Starting from the first equation of the BBGKY hierarchy, the
evolution equation for the one-particle distribution function $f_\alpha
(\mathbf{r}, \mathbf{v},t)$ of the free plasma charges is obtained as
\begin{equation} \label{kin}
m_{\alpha}\left(\partial_t + \mathbf{v}\partial_{\mathbf{r}}\right)
f_\alpha (\mathbf{r}, \mathbf{v},t) =  
 \sum_{\beta}\int{\left[\partial_{\mathbf{r}}\varphi_{\alpha\beta}(\mathbf{r},
\mathbf{r}^{\prime})\right]
\partial_{\mathbf{v}}f_{\alpha\beta}(\mathbf{r},\mathbf{v},\mathbf{r}^{\prime},
\mathbf{v}^{\prime},t)\:d\mathbf{r}^{\prime}d\mathbf{v}^{\prime}} \; ,
\end{equation}
where $\alpha, \beta$ label the particle species (e,i for electrons and ions,
respectively), $f_{\alpha\beta}(\mathbf{r},
\mathbf{v},\mathbf{r}^{\prime}, \mathbf{v}^{\prime},t)$ denotes the
two-particle distribution function for the corresponding particle species and
$\varphi_{\alpha\beta}=q_{\alpha}q_{\beta}/|{\bf{r}}-{\bf{r}}^{\prime}|$ is the
Coulomb interaction potential between the charges $q_{\alpha}$ and $q_{\beta}$.
Electron-electron correlations are very small during the plasma expansion,
since the electrons will quickly heat up due to three-body recombination and the
additional heating due to correlation effects is small in comparison
\cite{Rob02}. Hence,
we neglect electron-electron as well as electron-ion correlations, leaving
only IC as a possible influence on the plasma dynamics
beyond the mean-field level. On this level of approximation the ion kinetic
equation can be written as
\begin{equation} \label{ionkin}
m_{\rm{i}}\left(\partial_t + \mathbf{v}\partial_{\mathbf{r}}-\frac{\partial_{
\mathbf{r}}\bar{\varphi}}{m_{\rm i}}\partial_{\mathbf{v}}\right)
f_{\rm{i}} =  \\
 \int{\left(\partial_{\mathbf{r}}\varphi_{\rm{ii}}\right)\:
\partial_{\mathbf{v}}w_{\rm{ii}}(\mathbf{r},\mathbf{v},\mathbf{r}^{\prime},
\mathbf{v}^{\prime})\:d\mathbf{r}^{\prime}d\mathbf{v}^{\prime}}\;, 
\end{equation}
where the function
\begin{equation} \label{twodist}
w_{{\rm{ii}}}(\mathbf{r},\mathbf{v},\mathbf{r}^{\prime},\mathbf{v}^{\prime})=
f_{{\rm{ii}}}(\mathbf{r}, \mathbf{v},\mathbf{r}^{\prime},
\mathbf{v}^{\prime})-f_{{\rm{i}}}(\mathbf{r},
\mathbf{v})f_{{\rm{i}}}(\mathbf{r}^{\prime}, \mathbf{v}^{\prime})
\end{equation}
contains the contributions of  IC to the two-particle
distribution function and $\bar{\varphi}$ is the mean-field potential created
by all plasma charges. Since $m_{\rm e}/m_{\rm i} \ll 1$, the
relaxation timescale of the electrons is much smaller than the timescale of
the plasma expansion under typical experimental conditions \cite{Kil99}. Thus,
we may safely apply an adiabatic approximation for the electron distribution
function, assuming a local Maxwellian distribution
\begin{equation} \label{adi}
f_e(\mathbf{r},\mathbf{v}) \propto \exp \left(\frac{\bar{\varphi}(
\mathbf{r})}{k_{\rm B} T_{\rm e}} \right) \;
\exp \left(- \frac{\mathbf{v}^2}{2 k_{\rm B} T_{\rm e}} \right) \;\;,
\end{equation}
where $T_{\rm{e}}$ is the electron temperature. 
Eq.\ (\ref{adi}) together with a quasineutral approximation \cite{Dor98} allows
one to express the mean-field potential in terms of the ionic density
$\rho_{\rm{i}}=\int{f_{\rm{i}}d{\mathbf{v}}}$, resulting in
\begin{equation} \label{pot}
\partial_{\mathbf{r}} \bar{\varphi}=k_{\rm{B}}T_{\rm{e}} \frac{
\partial_{\mathbf{r}} \rho_{\rm{i}}}{\rho_{\rm{i}}} \; .
\end{equation}
Using Eq.\ (\ref{pot}), the following evolution equations
for the second moments of the ion distribution function are derived from
Eq.\ (\ref{ionkin})
\begin{subequations} \label{mom}
\begin{eqnarray} \label{moma}
\partial_t\left<r^2\right>&=&2\left<{\mathbf{r}}{\mathbf{v}}\right>
 \\
\label{momb}
\frac{m_{\rm{i}}}{2}\partial_t\left<{\mathbf{r}}{\mathbf{v}}\right>&=&
\frac{m_{\rm{i}}}{2}\left<v^2\right>+\frac{3}{2}k_{\rm{B}}T_{\rm{e}}
+\frac{1}{2}N_{\rm{i}}^{-1}\int{\rho_{\rm{i}}({\bf{r}}){\mathbf{r}}{
\mathbf{F}}_{\rm{ii}}({\mathbf{r}})\:d{\mathbf{r}}} \\ 
\label{momc}
\frac{m_{\rm{i}}}{2}\partial_t\left<v^2\right>&=&N_{\rm{i}}^{-1}
\left(k_{\rm{B}}T_{\rm{e}}
\int{\rho_{\rm{i}}\partial_{\mathrm{r}}{\mathbf{u}}\:d{\bf{r}}}
-\int{{\mathbf{v}}(\partial_{\mathbf{r}}\varphi_{\rm{ii}}})
w_{\rm{ii}}({\mathbf{r}},{\mathbf{v}},{\mathbf{r}}^{\prime},{
\mathbf{v}}^{\prime})\:d{\mathbf{r}}d{\mathbf{v}}d{\mathbf{r}}^{\prime}
d{\mathbf{v}}^{\prime}\right)
\end{eqnarray}
\end{subequations}
where $\left<r^2\right> = N_{\rm{i}}^{-1}\int r^2 f_{\rm i}({\mathbf{r}},
{\mathbf{v}}) d{\mathbf{r}} d{\mathbf{v}}$ etc.
The ``correlation force'' ${\mathbf{F}}_{\rm{ii}}({\mathbf{r}})$ is given by
\begin{equation} \label{cforce}
{\mathbf{F}}_{\rm{ii}}\left({\mathbf{r}}\right)=-\int{\left(\partial_{\mathbf{r
}}\varphi_{\rm{ii}}\right)\rho_{\rm{i}}({\bf{r}}^{\prime})g_{\rm{ii}}\left({
\mathbf{r}},{\mathbf{r}}^{\prime}\right)}\:d{\mathbf{r}}^{\prime} \;, 
\end{equation}
where the spatial correlation function $g_{\rm{ii}}$ is defined by
$\rho_{\rm i}({\mathbf{r}}) \rho_{\rm i}({\mathbf{r}}^{\prime})
g_{\rm ii}({\mathbf{r}},{\mathbf{r}}^{\prime}) \equiv \int w_{\rm{ii}}\left({
\mathbf{r}},{\mathbf{v}},{\mathbf{r}}^{\prime},{\mathbf{v}}^{\prime}\right)
\:d{\mathbf{v}}d{\mathbf{v}}^{\prime} $
and ${\mathbf{u}}({\mathbf{r}})=\int{{\mathbf{v}}f_{\rm{i}}}d{\mathbf{v}}$
is the hydrodynamical drift velocity of the plasma.
With the help of the second
kinetic ion equation of the BBGKY hierarchy, the last term on the right-hand
side of Eq.\ (\ref{momc}) can be written as
\begin{eqnarray} \label{cen}
\frac{1}{N_{\rm{i}}}\int{{\mathbf{v}}(\partial_{\mathbf{r}}\varphi_{\rm{ii}})
w_{\rm{ii}}\:d{
\mathbf{r}}d{\mathbf{v}}d{\mathbf{r}}^{\prime}d{\mathbf{v}}^{\prime}}&=&-
\frac{1}{2N_{\rm{i}}}\partial_{t}\int{\varphi_{\rm{ii}}w_{\rm{ii}}\:d{
\mathbf{r}}d{\mathbf{v}}d{\mathbf{r}}^{\prime}d{\mathbf{v}}^{\prime}}
\nonumber \\
&=&-\partial_{t}U_{\rm{ii}} \; ,
\end{eqnarray}
where 
\begin{equation} \label{Uc2}
U_{\rm{ii}} = \frac{1}{2 N_{\rm{i}}}\int \varphi_{\rm{ii}} \rho_{\rm{i}}({
\mathbf{r}})\rho_{\rm{i}}({\mathbf{r}}^{\prime})g_{\rm{ii}}({\mathbf{r}},{
\mathbf{r}}^{\prime}) \:d{\mathbf{r}}d{\mathbf{r}}^{\prime}
\end{equation}
is the average correlation energy per ion. Hence, Eq.\ (\ref{momc})
reflects energy conservation for the ion subsystem.
The evolution of the hydrodynamical velocity ${\mathbf{u}}$ is determined by
\begin{equation} \label{hyda}
m_{\rm{i}}\rho_{\rm{i}}\left[\partial_t{\mathbf{u}}+\left({\mathbf{u}}\cdot
\partial_{\mathbf{r}}\right){\mathbf{u}}\right]=-k_{\rm{B}}T_{\rm{e}}
\partial_{{\mathbf{r}}}\rho_{\rm{i}}-\partial_{\mathbf{r}}P_{\rm{th,i}}
-\rho_{\rm{i}}\mathbf{F_{\rm ii}}
\end{equation}
where $P_{\rm{th,i}}=\frac{m_{\rm{i}}}{3
N_{\rm{i}}}\int{({\mathbf{v}}-{\mathbf{u}})^2f_{\rm{i}}\:d{\mathbf{r}}d{
\mathbf{v}}}$ is the thermal ion pressure.
As shown in the appendix, in the framework of a local density approximation,
i.e.\ by assuming that $g_{\rm ii}$ only depends
on the distance $|\mathbf{r}-\mathbf{r}^{\prime}|$ and on the
densities at the two coordinates $\mathbf{r}$ and
$\mathbf{r}^{\prime}$, and that the ionic density $\rho_i$ varies slowly on the
lengthscale where $g$ is significantly different from zero,
the total correlation energy can be approximated by the well-known LDA
expression 
\begin{equation} \label{Uc}
U_{\rm ii}=N_{\rm{i}}^{-1}\int{u_{\rm{ii}}\rho_{\rm{i}}}\:d{\mathbf{r}}
\end{equation}
while the correlation force is found to be
\begin{equation}
\label{Fc}
\mathbf{F_{\rm{ii}}}
= -\frac{1}{3}\left(\frac{u_{\rm ii}}{\rho_{\rm i}}+\frac{\partial u_{\rm ii}}{
\partial \rho_{\rm i}}\right) \partial_{\mathbf{r}} \rho_{\rm{i}} \;\; ,
\end{equation}
where $u_{\rm ii}(\mathbf{r})$ is the correlation energy of a homogeneous plasma
of density $\rho_{\rm i}(\mathbf{r})$,
\begin{equation}
\label{uc}
u_{\rm ii}(\mathbf{r})=\frac{e^2}{2}\rho_{\rm i} (\mathbf{r}) \int
\frac{g_{\rm ii}(x; \rho_{\rm{i}})}{x} \, d{\mathbf{x}} \;\; .
\end{equation}
If  IC are neglected in Eq.\ (\ref{ionkin}), the kinetic equation exhibits the
following selfsimilar solution
\begin{eqnarray} \label{sol}
f_{\rm{i}}&\propto&\exp{\left(-\frac{r^2}{2\sigma^2}\right)}\exp{\left(-
\frac{m_{\rm{i}}\left({\bf{v}}-{\bf{u}}\right)^2}{2k_{\rm{B}}
T_{\rm{i}}}\right)} \nonumber \\
{\mathbf{u}}&=&\gamma{\mathbf{r}}
\end{eqnarray}
which corresponds to the initial state of the experiments under consideration.
As soon as IC are taken into account via the correlation
pressure $\rho_{\rm i} \bf{F}_{\rm{ii}}$ in Eq.\ (\ref{hyda}), however,
Eqs.\ (\ref{sol}) are no longer exact solutions of Eq.\ (\ref{ionkin}).
Using Eq.\ (\ref{Fc}), the last term on the right-hand side of Eq.\ (\ref{hyda})
can be rewritten as $\rho_{\rm{i}}{\bf{F}}_{\rm{ii}}=-\frac{1}{3}\frac{\partial
(u_{\rm{ii}}\rho_{\rm{i}})}{\partial \rho_{\rm{i}}}\partial_{\mathbf{r}}
\rho_{\rm i}$.
Interpreting this term as a local nonideal pressure, an equation for the
parameter $\gamma$ was derived in \cite{PPR03} by averaging
the differential equation for $|{\bf{u}}|/r$ obtained by inserting Eq.\
(\ref{sol}) into Eq.\ (\ref{hyda})
over the plasma volume. Obviously, this treatment is not unique.
Since, as discussed above, the ansatz (\ref{sol}) does not solve Eq.\
(\ref{hyda}) exactly, multiplying Eq.\ (\ref{hyda}) by different
functions of $r$ and averaging over the plasma volume will lead to slightly
different evolution equations for the parameter $\gamma$.
Here, supported by a comparison with our numerical results from the MD
simulations, to be discussed below, we assume as in \cite{PPR03} that the
functional form of the hydrodynamical
quantities of Eqs.\ (\ref{sol}) is not altered by the inclusion of IC, while
the dynamics of the parameters appearing in Eqs.\ (\ref{sol}) is determined from
the equations (\ref{mom}) for the moments of the distribution function.
Clearly, such an  approximation can not be {\sl a priori} justified. Hence, it
must be validated {\sl a posteriori} by comparison with more sophisticated
methods which do not rely on a reduction of the plasma description to a few
macroscopic parameters.

With this procedure, we arrive at the following set of equations
for the width $\sigma$ of the plasma cloud, its expansion velocity $\gamma
\mathbf{r}$ as well as ionic and electronic temperature $T_{\rm i}$ and
$T_{\rm e}$:
\begin{subequations}\label{par}
\begin{eqnarray} 
\label{para}
\partial_{t}\sigma^2&=&2\gamma\sigma^2  \\
\label{parb}
\partial_t\gamma&=&\frac{k_{\rm{B}}T_{\rm{e}}+k_{\rm{B}}T_{\rm{i}}+\frac{1}{3}
U_{\rm{ii}}}{m_{\rm{i}}\sigma^2}-\gamma^2
\\
\label{parc}
\partial_t k_{\rm{B}}T_{\rm{i}}&=&-2\gamma
k_{\rm{B}}T_{\rm{i}}-\frac{2}{3}\gamma U_{\rm{ii}}-\frac{2}{3}\partial_t
U_{\rm{ii}} \\
\label{pard}
\partial_t k_{\rm{B}}T_{\rm{e}}&=&-2\gamma k_{\rm{B}}T_{\rm{e}} \; .
\end{eqnarray}
\end{subequations}
The last equation (\ref{pard}) has been derived from the electron kinetic
equation by making use of the quasineutrality condition.
The set of Eqs.\ (\ref{par}) slightly differs from that presented in
\cite{PPR03} where $W_{\rm{c}}=\frac{1}{3N_{\rm{i}}}\int \rho_{\rm{i}}\frac{
\partial
(u_{\rm{ii}}\rho_{\rm{i}})}{\partial \rho_{\rm{i}}}\:d{\bf{r}}$ had been
used instead of $U_{\rm{ii}}$ in Eq.\ (\ref{parb}). A
comparison with our numerical MD results shows that Eqs.\ (\ref{par}) yield a
slightly better quantitative agreement, while the principal influence of 
IC on the plasma dynamics, which has been partly discussed in
\cite{PPR03}, is the same.
Eqs.\ (\ref{par}) provide a transparent physical
picture of the expansion dynamics. First, Eq.\ (\ref{pard}) together with
(\ref{para}) reflects the
adiabatic cooling of the electron gas, i.e.\ $T_{\rm{e}}\sigma^2={\rm{const}}$.
The ion temperature, on the other hand, is not only affected by the adiabatic
cooling, expressed by the first term in Eq.\ (\ref{parc}), but also changes due
to the development of IC, which is taken into account by the last term in
Eq.\ (\ref{parc}). Furthermore, these correlations reduce the ion-ion
interaction and therefore lead to an effective negative acceleration,
expressed by the $U_{\rm{ii}}/3$-term in Eq.\ (\ref{parb}), in
addition to the ideal thermal pressure.
This contribution, which corresponds to the average nonideal pressure
known from homogeneous systems \cite{Ich82,Dub99}, also leads to an effective
potential in which the ions
move. As they expand in this potential, the thermal energy changes due to
energy conservation, as expressed by the second term on the right-hand side of
Eq.\ (\ref{parc}). Finally, combining eqs.(\ref{par}) yields a second integral
of motion, namely the total energy of the plasma
\begin{equation}
E_{\rm{tot}}=\frac{3}{2}\left(k_{\rm{B}}T_{\rm{e}}+k_{\rm{B}}T_{\rm{i}}\right)
+\frac{3}{2}m_{\rm{i}}\gamma^2\sigma^2+U_{\rm{ii}}\;.
\end{equation}
Although the set of equations (\ref{par}) determines the time evolution of all
relevant macroscopic plasma parameters, namely its width, expansion velocity,
electron and ion temperature,  it is not a closed set since an evolution
equation for the correlation energy $U_{\rm{ii}}$ which enters Eq.\
(\ref{parc}) is missing. Initially, the plasma is completely uncorrelated, so
that $U_{\rm ii} (t=0) = 0$. However, the initial state corresponds to a
non-equilibrium
situation, and the plasma will relax towards thermodynamic equilibrium,
thereby building up correlations. A precise description of this
relaxation process in the framework of a kinetic theory is rather complicated
and requires a considerable numerical effort \cite{Sem99}. We therefore employ a
linear approximation for the relaxation of the two-particle correlation
function, the so-called correlation-time approximation \cite{Bon96a},
\begin{equation} \label{CTA}
\frac{dw_{\rm ii}({\bf{r}},{\bf{v}},{\bf{r}}^{\prime},{\bf{v}}^{\prime};t)}{dt}
\approx-\frac{w_{\rm ii}({\bf{r}},{\bf{v}},{\bf{r}}^{\prime},{\bf{v}}^{\prime};
t)-w_{\rm ii}^{\rm{eq}}({\bf{r}},{\bf{r}}^{\prime};t)}{\tau_{{\rm{corr}}}} \; .
\end{equation}   
Here, $\tau_{{\rm{corr}}}$ \cite{Bon96a, Bon96b} is the characteristic
timescale for the relaxation
of particle correlations and $w_{{\rm{ii}}}^{{\rm{eq}}}$ is the equilibrium
pair correlation function, which in our case still depends on time via the
evolving one-particle distribution function since the plasma is freely
expanding. As shown in \cite{Mor98} the correlation time $\tau_{\rm{corr}}$ can
be well estimated by the inverse ionic plasma frequency. Hence, in our
calculations we set $\tau_{\rm{corr}}=\omega_{\rm{p,i}}^{-1}=\sqrt{m_{\rm{i}}/
(4\pi e^2\bar{\rho}_{\rm{i}})}$, where
$\bar{\rho}_{\rm{i}}=N_{\rm{i}}/(4\pi\sigma^2)^{3/2}$ is the average ionic
density of the plasma. 
Such a linear approximation is good only for small
deviations of $w_{{\rm{ii}}}$ from its equilibrium form. Clearly, this is not
the case in the initial stage of the gas evolution. However, after the initial
phase of correlation heating the system stays very close to its slowly
changing local equilibrium, and one may expect Eq.~(\ref{CTA}) to yield good
results. Under the same conditions that lead to Eqs.\ (\ref{Uc}) and (\ref{Fc}),
one easily verifies that Eq.\ (\ref{CTA}) leads to
\begin{equation} \label{UCTA}
\partial_t U_{\rm{ii}}\approx-\frac{U_{\rm{ii}}-U_{\rm{ii}}^{\rm{eq}}}{\tau_{{
\rm{corr}}}},
\end{equation}  
where $U_{\rm{ii}}^{\rm{eq}}=N_{\rm{i}}^{-1}\int{\rho_{\rm{i}}u_{\rm{ii}}^{
\rm{eq}} d{\mathbf{r}}}$
and $u_{\rm{ii}}^{\rm{eq}}({\mathbf{r}})$ is the correlation energy
per particle of a homogeneous one-component plasma in local equilibrium.
This quantity has been studied intensively in the past, and approximate
analytical formulae are available in the literature
\cite{Ich82,Dub99}. Here, we adopt the interpolation formula from \cite{Cha98}
\begin{equation} \label{eos}
u_{\rm{ii}}^{\rm{eq}}({\mathbf{r}})=k_{\rm B} T_{\rm i} \Gamma^{3/2}\left(
\frac{A_1}{\sqrt{A_2+ \Gamma}}+ \frac{A_3}{1+\Gamma}\right) \;,
\end{equation}
with $A_1=-0.9052$, $A_2=0.6322$ and $A_3=-\sqrt{3}/2-A_1/\sqrt{A_2}$,
which yields an accurate interpolation between the low-$\Gamma$ Abe limit and
the high-$\Gamma$ behavior obtained by Monte Carlo and MD simulations.
It should be noted that in the present situation $u_{\rm ii}$ depends on time
since the plasma expands. Hence, $\Gamma$ and with it the thermodynamical
equilibrium change in time.

The set of equations (\ref{par}) describes the evolution of the plasma
part of the system, i.e.\ a system of $N_{\rm i}$ ions and electrons.
Due to ionization and recombination events occurring during the plasma
expansion (discussed in detail in the following subsection), this number
$N_{\rm i}$, and hence also the total mass $M = N_{\rm i} m_{\rm atom}$, is not
constant
over the course of the evolution. However, such a treatment completely
neglects the influence of the bound Rydberg atoms on the dynamics. One may
argue that they do not influence the plasma evolution since they do not
interact with the ions or electrons by Coulomb interaction. On the other hand,
a Rydberg atom may carry a significant amount of kinetic energy, gained from the
acceleration by the electron pressure before its formation by three-body
recombination. In a simple approximation, we
assume equal hydrodynamical velocities and density profiles for the
ions and atoms, in order to account for this effect. This implies that the
expansion of the neutral Rydberg atoms can be taken into account by replacing
the mass
$N_{\rm{i}}m_{\rm{i}}$ of the ions by the mass of the {\em total} system
$(N_{\rm{i}}+N_{\rm{a}})m_{\rm{i}}$, where $N_{\rm{a}}$ is the number of
atoms. We therefore replace the ion mass $m_{\rm{i}}$ by an effective mass
$(1+N_{\rm{a}}/N_{\rm{i}})m_{\rm{i}}$ in Eq.\ (\ref{parb}).
The quality of this approximation can, of course, also be checked by
comparison with the H-MD description, see below.
\subsection{Ionization and Recombination}
\label{zb}
As demonstrated in \cite{Rob02}, a satisfactory description of
the dynamics of an ultracold plasma can be achieved by combining a
hydrodynamic treatment of the plasma evolution with rate
equations accounting for inelastic collisions between the plasma particles and
Rydberg atoms. 
The rate equation for the change of density
of Rydberg atoms in a state with principal quantum number $n$ reads
\begin{equation}
\label{rate}
\dot{\rho}_{\rm a}(n)=\rho_{\rm e}\sum_{p}\left[K(p,n)\rho_{\rm a}(p)- K(n,p)
\rho_{\rm a}(n) \right]+\rho_{\rm e}\left[R(n)\rho_{\rm e}\rho_{\rm i}-I(n) \rho_{\rm a}(n)\right]
\;, 
\end{equation}
where \(K(p,n)\) is the rate coefficient for electron impact (de)excitation
from level $p$ to level $n$, and $R(n)$ and $I(n)$ describe
three-body recombination into and electron-impact ionization from level $n$,
respectively.
The rate coefficients $K$, $R$ and $I$ have been taken from the classic
work of Mansbach and Keck \cite{Man69}. Additional processes, such as, e.g.,
ionization by black-body radiation or from dipolar atom-atom interactions,
are easily included in Eq.\ (\ref{rate}) if the corresponding rates are
available. Such processes are essential for a description of the early stages
of the evolution of a system starting with a Rydberg gas
\cite{Rob00,Eyl00,Gou03}, but are of minor importance in situations starting
from a pure plasma.

In this framework, the evolution of the system is obtained by solving
Eqs.\ (\ref{para})-(\ref{parc}) together with Eq.\ (\ref{rate}) while the
electron temperature is now obtained from the modified energy conservation
relation $E_{\rm{tot}}+E_{\rm{a}}={\rm{const.}}$ instead of Eq.\ (\ref{pard}),
where
$E_{\rm{a}}=-{\cal{R}}\sum_{n}{N_{\rm{a}}n^{-2}}$ is the total energy of the
Rydberg atoms and ${\cal{R}}=13.6\:{\rm{eV}}$.
\subsection{Hybrid molecular dynamics treatment}
\label{zc}
As we will show in section III, the kinetic description of the previous
subsections is able to describe the plasma dynamics to a surprisingly large
extent. However, one of the main motivations of this work is the study of the
role of IC, which are incorporated in the model only in an
approximate way. To assess their influence on the dynamics reliably,
a more sophisticated approach is required, e.g.\ molecular dynamics simulations
which fully incorporate the ionic interactions. However, a full MD simulation of
both, electrons and ions, is computationally very demanding, and only the very
early stage of the system evolution can be described in this way \cite{Kuz02}.
On the other hand, as argued above, electronic correlations are not
important for the plasma dynamics, so that only IC have to
be accounted for in full while the influence of the electrons on the dynamics
may be treated on a mean-field level. Moreover, we have seen that the
timescale of equilibration of the electronic subsystem is orders of magnitude
shorter than that of the ionic subsystem and the timescale of the plasma
expansion. This observation led us to use an adiabatic approximation in
subsection \ref{za}, where
the electrons are assumed to equilibrate instantaneously, assuming a Maxwellian
velocity distribution with a well-defined temperature and a spatial profile
determined from the total mean-field potential of the
plasma charges. The clear separation of timescales suggests that this
adiabatic approximation is well justified, hence we will keep it in the
following. Consequently, we have developed a hybrid approach where the electrons
are treated on a hydrodynamical level as in the kinetic description above,
while the ions are propagated individually with their mutual 
interaction and the influence of the electrons
on the ions enters via the electronic mean-field potential. This hybrid approach
permits the use of much larger timesteps in the propagation of the system,
since the electronic dynamics needs not to be followed in detail but only the
ionic motion
has to be resolved in time. Consequently, the evolution of the system can be
followed over the experimental timescales. 
Furthermore, the approximate treatment of IC in the kinetic
model of subsection \ref{za} can be tested.
Finally, beyond the scope of the present work, we have shown \cite{PPR04} that
the present H-MD approach can describe situations  where the ionic plasma
component is so strongly coupled that crystallization of the ions sets in. Such
a scenario is clearly beyond the capabilities of a kinetic approach.

As discussed above, the
electrons are still treated as a fluid, while we lift the
quasineutral approximation by calculating the resulting mean-field potential
from the Poisson equation
\begin{equation} \label{poisson}
\Delta \bar{\varphi}=4\pi e^2\left(\rho_{\rm{e}}-\rho_{\rm{i}}\right).
\end{equation}
However, using Eq.\ (\ref{adi}) poses a conceptual difficulty \cite{Rob03}
since the
mean-field potential approaches a finite value at large distances and therefore
leads to a non-normalizable electron density. This problem, which has been
discussed for a long time in an astrophysical context
\cite{Cha43}, reflects the fact that a substantial fraction of the
electrons indeed escapes the finite potential barrier at long times during the
relaxation process until the total kinetic energy of all electrons is less
than the height of the potential well. On the timescales under
consideration, however, typically only a small amount of the electrons escapes the
plasma volume, until the resulting charge imbalance becomes large enough to trap the
remaining electrons, which quickly reach a quasi-steady state forming a
temporarily quasineutral plasma in the central region. We account for this
electron loss by determining the fraction of trapped electrons from the
results of Ref.~\cite{Kil99}. 

The corresponding steady-state distribution, derived for the
study of globular clusters, is of the form \cite{Kin66}
\begin{equation} \label{king1}
\rho_{\rm{e}}\propto \exp{\left(\frac{\bar{\varphi}}{k_{\rm{B}}T_{\rm{e}}}
\right)}\int_{0}^{\chi}\exp{\left(-x\right)}x^{3/2}\:dx,
\end{equation}
where $\chi=m_{\rm{e}}v_{\rm{esc}}^2/(2k_{\rm{B}}T_{\rm{e}})$ with the velocity 
$v_{\rm{esc}}(\bf{r})$ necessary to escape from a given position
in the plasma. In the present case, the potential can have a non-monotonous
radial space dependence and the escape velocity has to be defined as
\begin{equation} \label{king2}
\frac{m_{\rm{e}}}{2}v_{\rm{esc}}^2\left(r\right)=\max_{r^{\prime}\geq r}{\left[
\bar{\varphi}\left(r^{\prime}\right)-\bar{\varphi}\left(r\right)\right]},
\end{equation}
in contrast to astrophysical problems where one only has a
single sign of ``charge'' and $m_{\rm{e}}v_{\rm{esc}}^2/2=-\bar{\varphi}$
\cite{Kin66}. For a given electron temperature $T_{\rm{e}}$ and ion density
$\rho_{\rm i}$ the electron density is found by numerical iteration of Eqs.\
(\ref{poisson}), (\ref{king1}) and (\ref{king2}) until selfconsistency is
reached. 

Knowledge of the electron density then permits
propagation of the ions in the electron mean-field
$\Delta\bar{\varphi}_{\rm{e}}=4\pi e^2\rho_{\rm{e}}$ and the full interaction
potential of the remaining ions,
\begin{equation} \label{eom}
m_{\rm{i}}\ddot{\bf{r}}_{j}=\partial_{\bf{r}_{j}}\bar{\varphi}_{\rm{e}}+e^2
\sum_{k}\frac{{\bf{r}}_{j}-{\bf{r}}_{k}}{\left|{\bf{r}}_{j}-{\bf{r}}_{k}
\right|^3} \;.
\end{equation}
The numerical solution of the ion equations of motion represents the most time
consuming part of the plasma propagation. In general, for $N$ propagated
particles, the corresponding numerical effort scales with $N^{2}$
rendering a treatment of large particle numbers difficult. In order
to simulate particle numbers relevant to the experiments, we have adapted a
hierarchical treecode originally designed for astrophysical problems,
first described in \cite{Bar86}. This method provides a numerically
exact solution of the ion equations of motion Eq.\ (\ref{eom}), while the
numerical effort grows only as $N\ln N$ with increasing  $N$. 
More details about the numerical procedure can be found, e.g., in
\cite{Bar90}.

In the framework of the kinetic model introduced in section \ref{za}, the
influence of IC on the system evolution can be singled out by
comparison with the solution of the corresponding equations with
$U_{\rm{ii}}\equiv 0$. In order to make an analogous comparison also for the MD
simulations, we have performed calculations propagating the ions in the
mean-field potential created by all charges. Technically, the mean-field
potential is represented using a test-particle method, widely used for various
problems in plasma physics (see, e.g., \cite{Bir95}).
\section{Results and Discussion}
We will discuss the evolution of a plasma initially consisting of $N_{\rm e} =
37500$ electrons and
$N_{\rm i} = 40000$ ions with an average density of 10$^9$ cm$^{-3}$
at a rather low electronic kinetic energy $E_{\rm e} = 3 k_{\rm B} T_{\rm e}
/2 = 20$ K, comparing the results from the kinetic model and our MD simulation.
Thereby, we put special emphasis on the role of IC.

\subsection{Global aspects of plasma expansion and recombination}
\label{GlobalResults}

\begin{figure}[bt]
\centerline{\psfig{figure=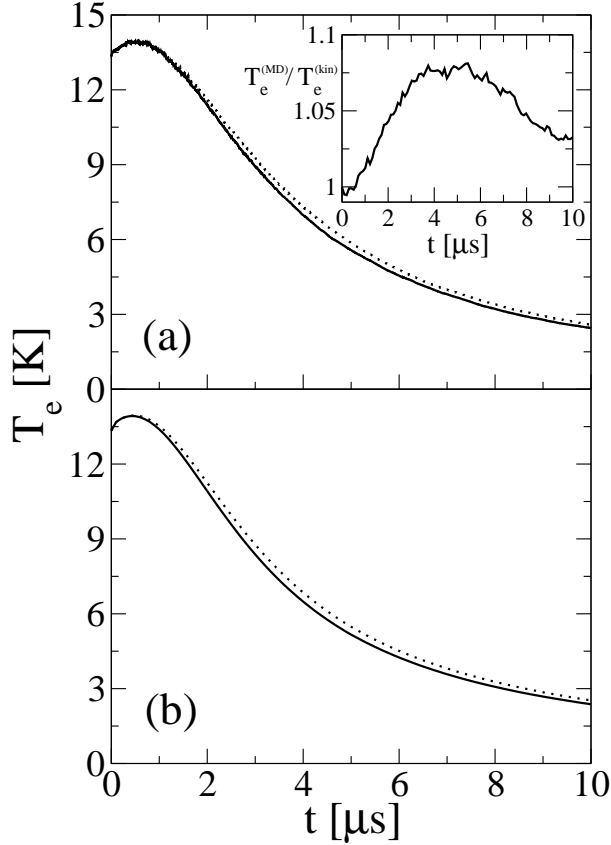,width=8cm}}
\caption{\label{fig1}
Electronic temperature $T_{\rm e}(t)$ for an expanding plasma of $40000$ ions
with an initial average density of $10^9$cm$^{-3}$ and an initial electron
kinetic energy of $20\:$K, obtained from the H-MD simulation (a) and the
kinetic model (b), with (solid) and without (dotted) the inclusion of
IC. The inset shows the ratio of the electron temperatures obtained from the
H-MD simulation and the kinetic model.}
\end{figure}

The general macroscopic behavior of the system has been described before in
several publications, experimentally as well as theoretically
\cite{Kul00,Kil01, Rob02,Rob03}. The
plasma cloud slowly expands due to the thermal pressure of the electrons,
leading to adiabatic cooling of the electrons as well as partial recombination
into bound states (figures \ref{fig1} and \ref{fig2}).
The amount of recombination and its influence strongly depends on the initial
electron temperature and density. If the electrons are too hot (about $50\:$ K
for typical experimental densities of $10^9\:$ cm$^{-3}$), recombination is
strongly suppressed and the system dynamics is well described by the results
of \cite{Dor98} obtained for the collisionless plasma expansion \cite{Kul00}.

\subsubsection{Temporal evolution of the electronic temperature}
For the lower electron temperatures considered here, as can be seen in Fig.\
\ref{fig1}, there is an initial increase of the
electron temperature due to electron heating by three-body recombination and
subsequent deexcitation of the formed Rydberg atoms. At low initial electron
energies this heating drastically increases the electron temperature and thus
accelerates the plasma expansion \cite{Rob02}, which
explains the enhanced expansion velocity observed in \cite{Kul00}. In contrast
to this recombination heating of the electrons, the inclusion of IC
 only slightly changes the expansion dynamics, as seen in Fig.\
\ref{fig1} by comparing the solid and dotted lines. As shown in the inset of
Fig.\ \ref{fig1}a, the electron temperature obtained from the H-MD
simulation and the kinetic model differ by at most $8$\% during the first few
microseconds of the plasma expansion, while the agreement becomes even better at
later times. Moreover, the faster decrease of the electron temperature due to
the inclusion of IC, predicted by the particle simulations, is quantitatively
reproduced by the much simpler kinetic model.

Hence, the simple evolution equations (\ref{par}) are sufficient to clarify
 the role of IC in the expansion
dynamics. According to Eq.\ (\ref{parc}), the development of IC
quickly heats up the plasma ions to roughly $-\frac{2}{3}U_{\rm{ii}}$
since the expansion of the plasma is still negligible during this initial stage.
Thereby,  the negative correlation
energy term $\frac 13U_{\rm ii}$ in Eq.\ (\ref{parb}) is overcompensated,
leading to a faster expansion of the plasma. As a consequence of the quicker
expansion, the electron temperature decreases somewhat faster than without the
inclusion of IC. With Eq.\ (\ref{parb}), the importance of this effect can be
estimated by comparing the thermal electron energy $k_{\rm B} T_{\rm e}$ to
 the net ion contribution $-\frac{1}{3}U_{\rm{ii}}$ in the numerator
of the first term on the right-hand side of Eq.\ (\ref{parb}). Estimating the
correlation energy by $e^2/a$, it follows that the
total pressure driving the plasma expansion is enhanced by a factor of roughly
$1+\Gamma_{\rm e}/3$, which only slightly changes the expansion dynamics since
the electrons are known to be weakly coupled over the whole observation time
\cite{Rob02}.
 
\subsubsection{Formation of Rydberg atoms in time}
\begin{figure}[bt]
\centerline{\psfig{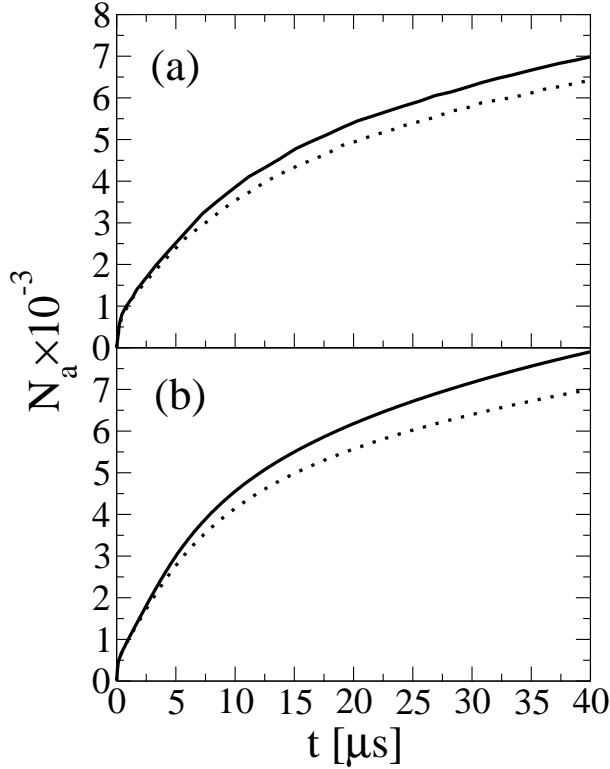}}
\caption{\label{fig2}
Number $N_a(t)$ of recombined atoms obtained from the H-MD simulation (a)
and the kinetic model (b) for the same parameters as in Fig.\
\ref{fig1}. The solid line shows the result taking into account the
IC while the dotted line is obtained from the mean-field treatment.}
\end{figure}

The number of recombined atoms is influenced more strongly by IC
(Fig.~\ref{fig2}). During the evolution of
the system, Rydberg atoms are constantly formed by three-body recombination
and re-ionized by the free electrons in the plasma. As shown in Fig.\
\ref{fig2}a, for the current set of parameters about $7000$ Rydberg atoms are
present in the system after $40\:\mu$s, while the kinetic model yields about
$7900$ atoms at the same instant of time (Fig.\ \ref{fig2}b). 
This number is small compared to the size of the whole system, nevertheless it
is large enough that the recombined atoms can be detected in an experiment, and
corresponding curves have indeed been obtained experimentally \cite{Kil01}. Due
to the strong temperature dependence of the total three-body recombination rate,
which is proportional to $T_{\rm e}^{-9/2}$ \cite{Man69}, the slight decrease of
the electron temperature due to the faster expansion, caused by the correlation
heating of the ions,
considerably affects the recombination behavior of the plasma. While there is
an overall shift between the atom number obtained from the particle simulations
and the kinetic model, both the
kinetic model and the H-MD simulation yield an increase of the atom
number of about $10\%$  at $t=40\:\mu$s (Fig.\ \ref{fig2}), compared to a
mean-field treatment of the ion dynamics. Thus, the
H-MD simulation corroborates our previous findings \cite{PPR03}.

\begin{figure}[t]
\centerline{\psfig{figure=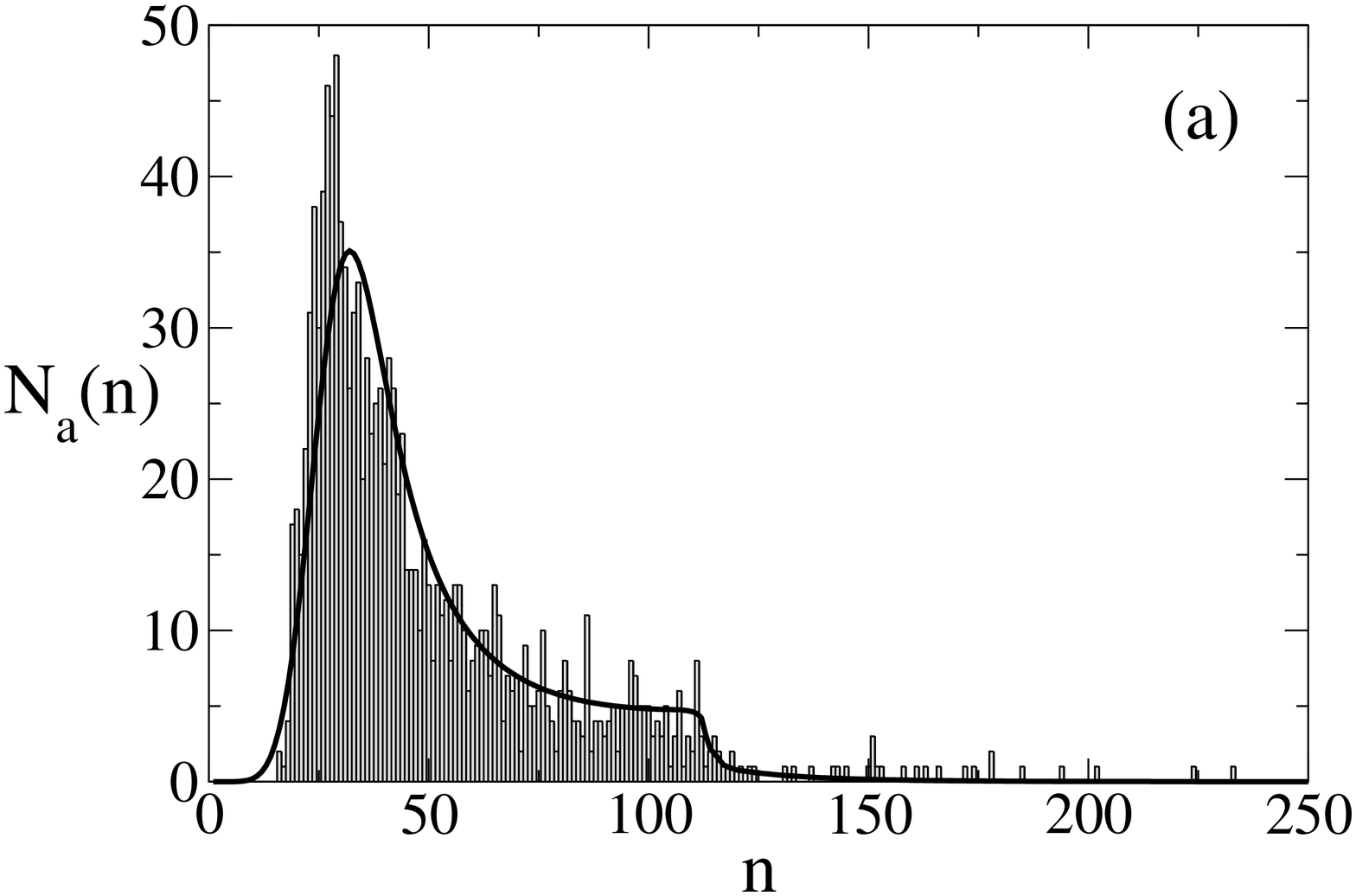,width=8cm}}
\centerline{\psfig{figure=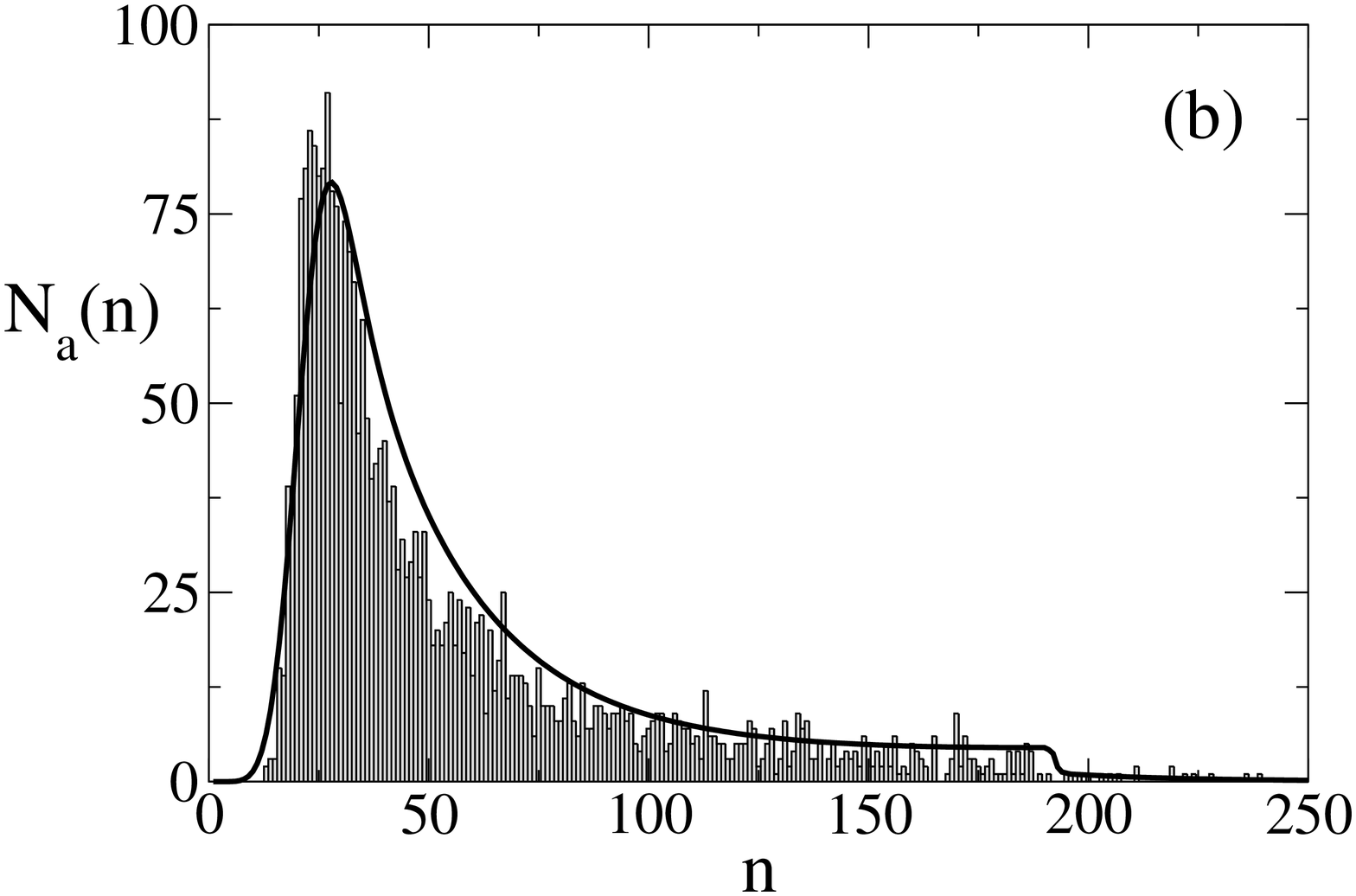,width=8cm}}
\centerline{\psfig{figure=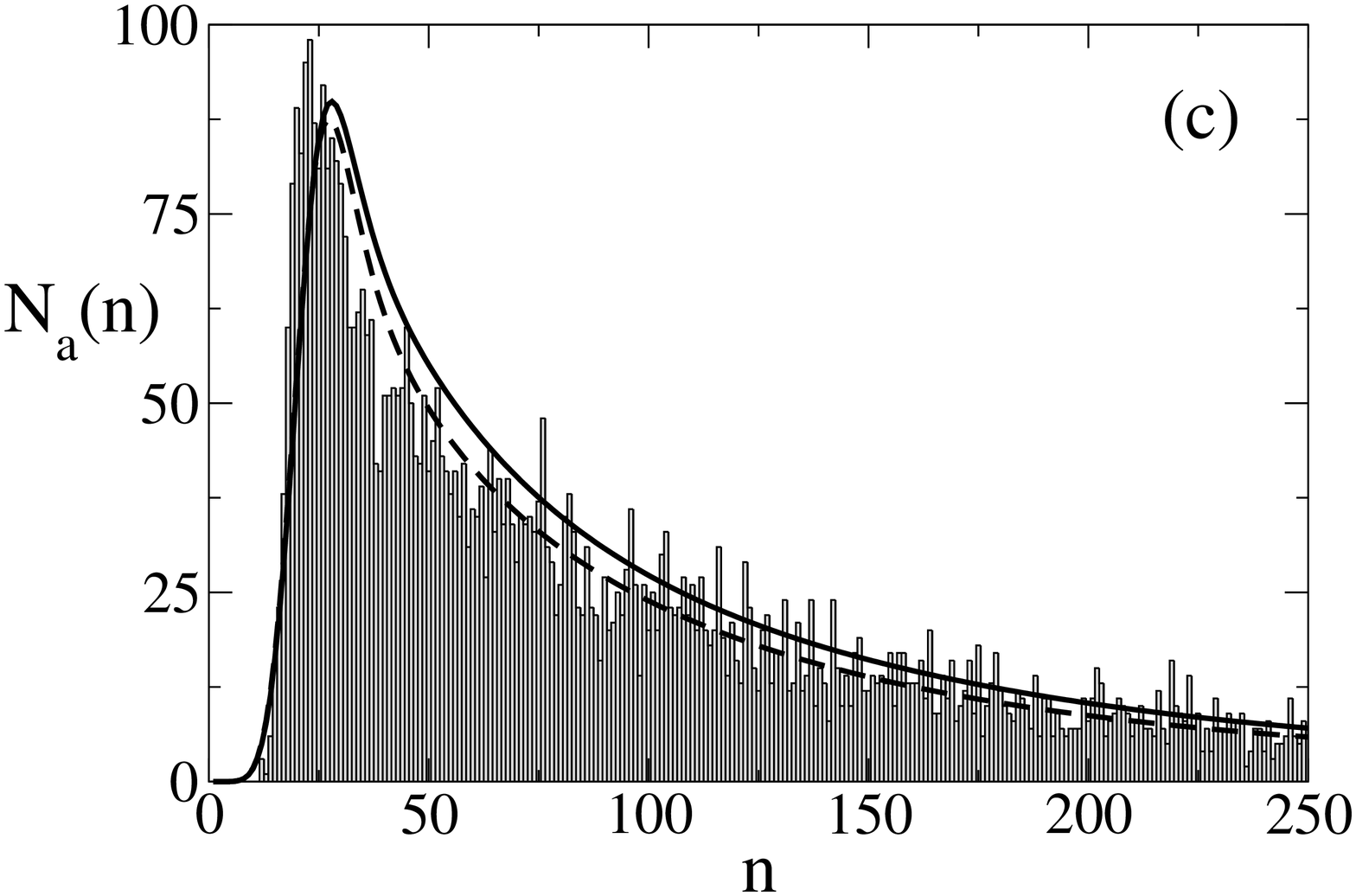,width=8cm}}
\caption{\label{fig3}
Population of bound Rydberg states with principal quantum number $n$, after
$t=1.5\:\mu$s (a), $t=6\:\mu$s (b) and $t=40\:\mu$s (c). The vertical bars
represent the
H-MD calculation, the solid line the kinetic model. The dashed curve
in (c) shows the kinetic model neglecting IC. Initial-state parameters are
the same as in Fig.\ \ref{fig1}.}
\end{figure}
Additional insight into the recombination process can be gained from a closer
look at the distribution of bound Rydberg states. Figure \ref{fig3} shows
the population of levels with principal quantum number $n$ for three different
times, corresponding to different stages of the plasma expansion. Initially, 
Rydberg states of moderate excitation are populated, due to a relatively high
electron temperature (Fig.\ \ref{fig3}a). At later times, higher excited bound
states are formed in the course of the plasma expansion (Figs.~\ref{fig3}b and
\ref{fig3}c), since the maximum principal quantum number for recombination 
$n_{\rm{max}}=\sqrt{{\cal{R}}/(2k_{\rm{B}}T_{\rm{e}})}$ \cite{Man69} increases
as the electron temperature drops down.
Moreover, the deeply bound states formed at earlier times are also not
subject to electron-impact excitation and deexcitation anymore since the
thermal velocity of the impacting electrons has become too small. Thus, as
becomes apparent by comparing Fig.\ \ref{fig3}b with \ref{fig3}c, the
deeply bound states ($n \alt 30$) remain basically untouched, while higher and
higher states ``freeze out'' as the plasma expands. As may be anticipated from 
Fig.~\ref{fig2}, IC mainly affect the later stages of the plasma
evolution. Hence, the inclusion of IC alters the
population of these higher lying states, as shown in Fig.\ \ref{fig3}c. Since
these states have small binding energy, they contribute little to the total
kinetic energy of the plasma subsystem. This is the reason why
the effect of IC is visible in the distribution of  Rydberg states,
but not in the macroscopic expansion dynamics of the plasma, reflected, e.g., by
the asymptotic expansion velocity measured in \cite{Kul00}. 
\subsection{Spatially resolved plasma expansion and relaxation}
While the time evolution of global, i.e.\ space-averaged, observables of the
plasma is very well described by the kinetic model, one may expect
discrepancies compared to the MD simulations when looking into the spatially
resolved plasma dynamics. We will assess these discrepancies quantitatively in
the following.
\begin{figure}[bt]
\centerline{\psfig{figure=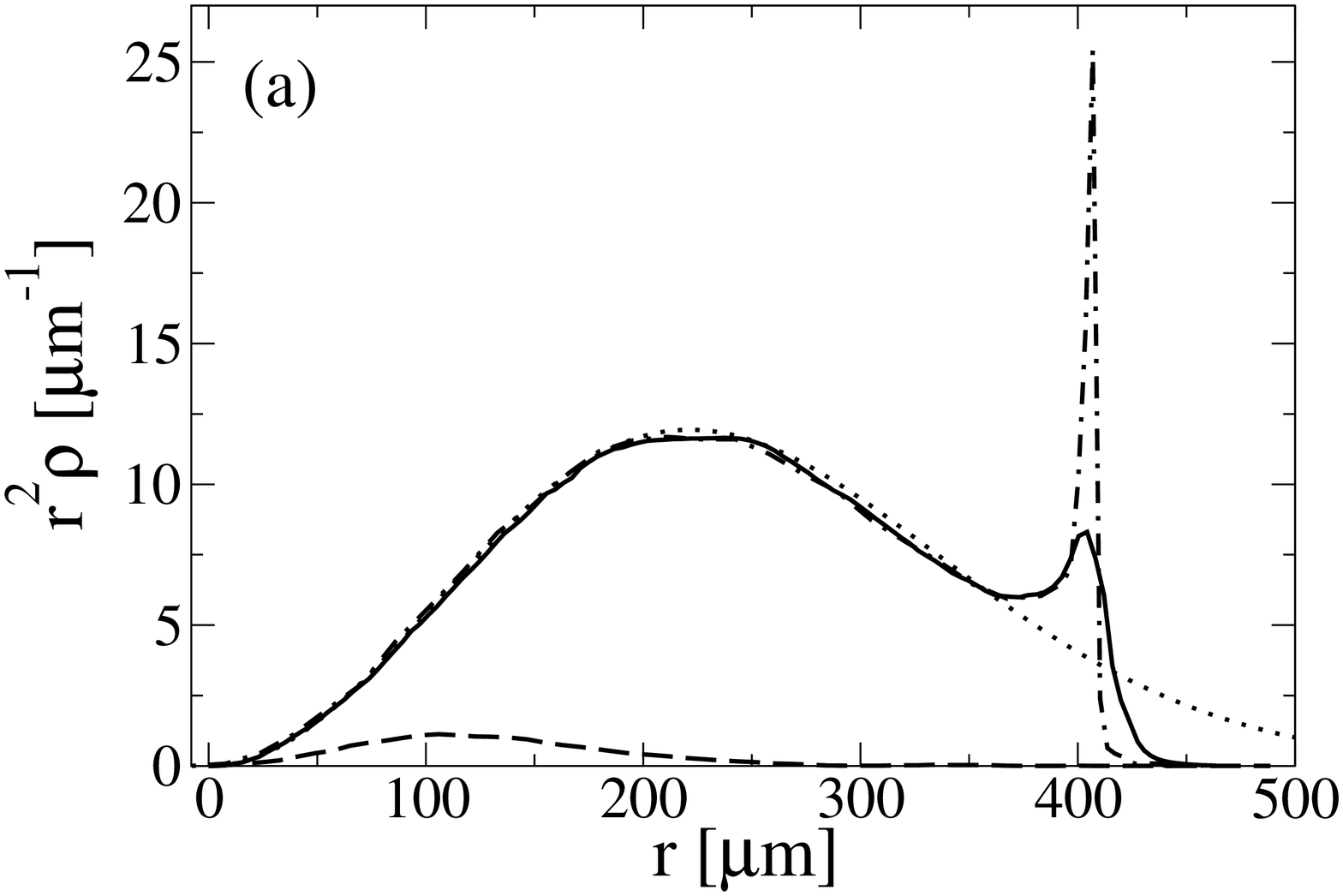,width=8cm}}
\centerline{\psfig{figure=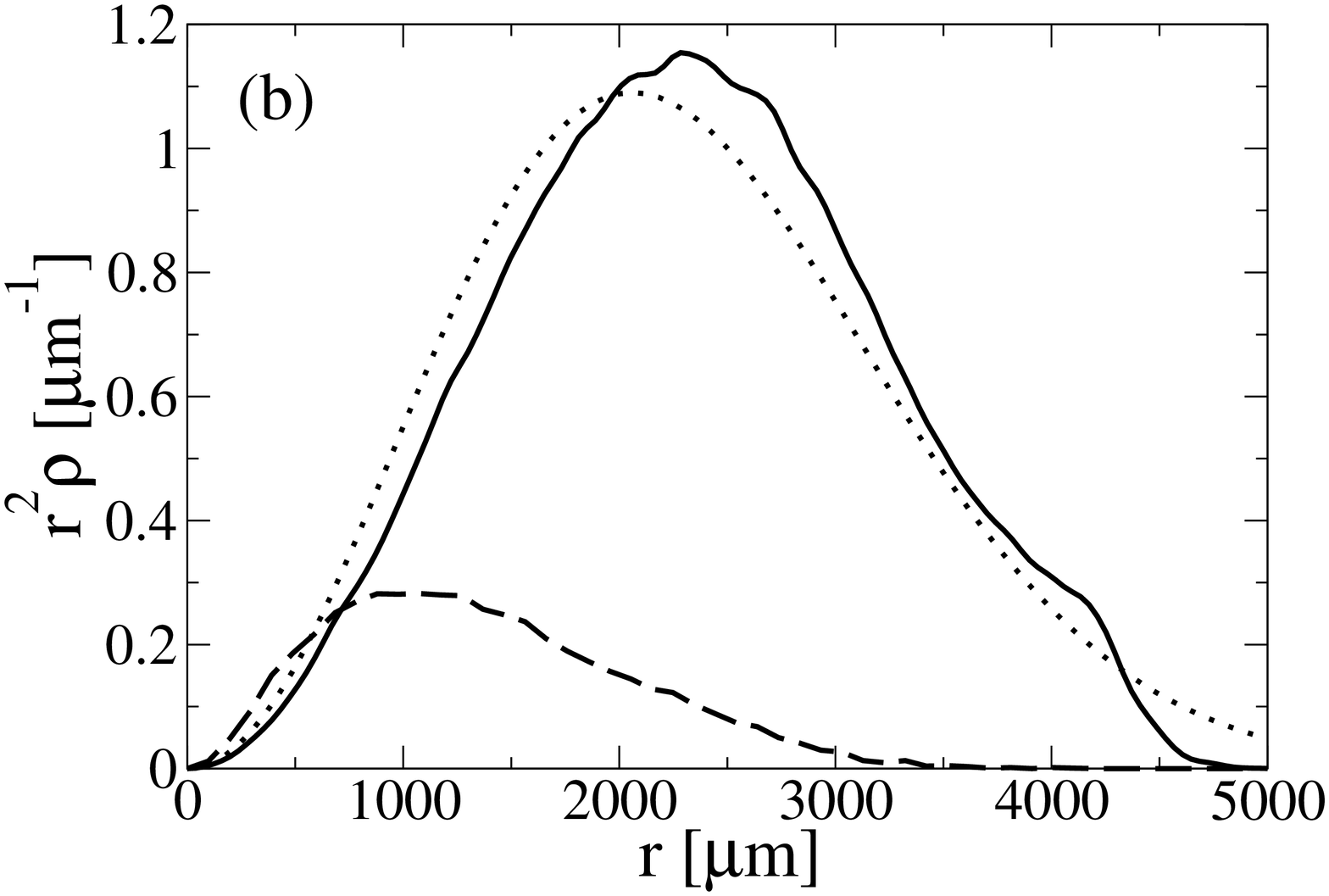,width=8cm}}
\caption{\label{fig4}
Spatial densities $\rho_{\rm i}$ (solid) and $\rho_{\rm a}$ (dashed)
of the ions and atoms, respectively, at $t=3\:\mu$s (a) and
$t=31.3\:\mu$s (b), compared to the Gaussian profile assumed for the kinetic
model (dotted). Additionally, $\rho_{\rm i}$ obtained from the particle
simulation using the mean-field interaction only is shown as the dot-dashed
line in (a). Initial-state parameters are the same as in Fig.\ \ref{fig1}.}
\end{figure}

\begin{figure}[tb]
\centerline{\psfig{figure=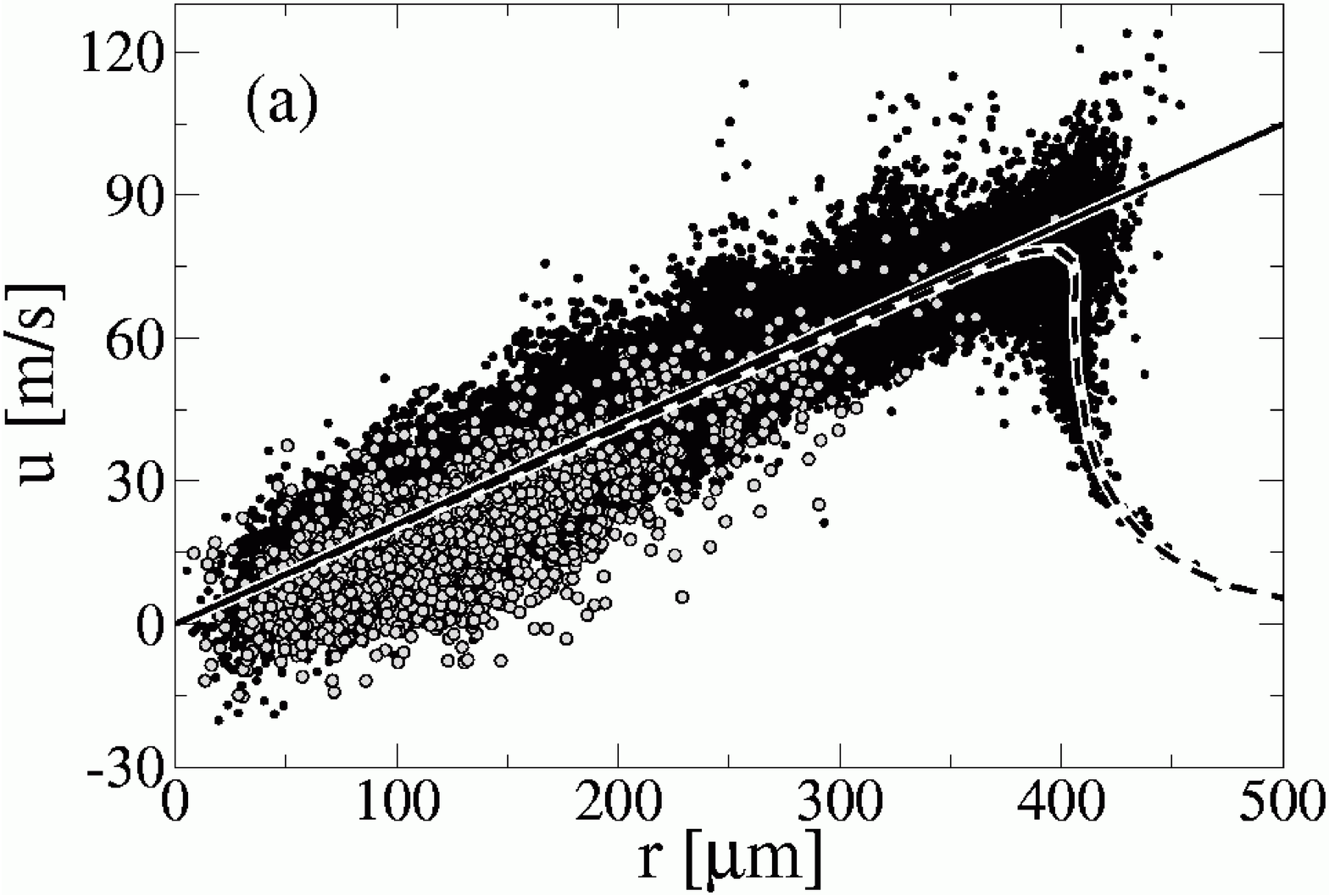,width=8cm}}
\centerline{\psfig{figure=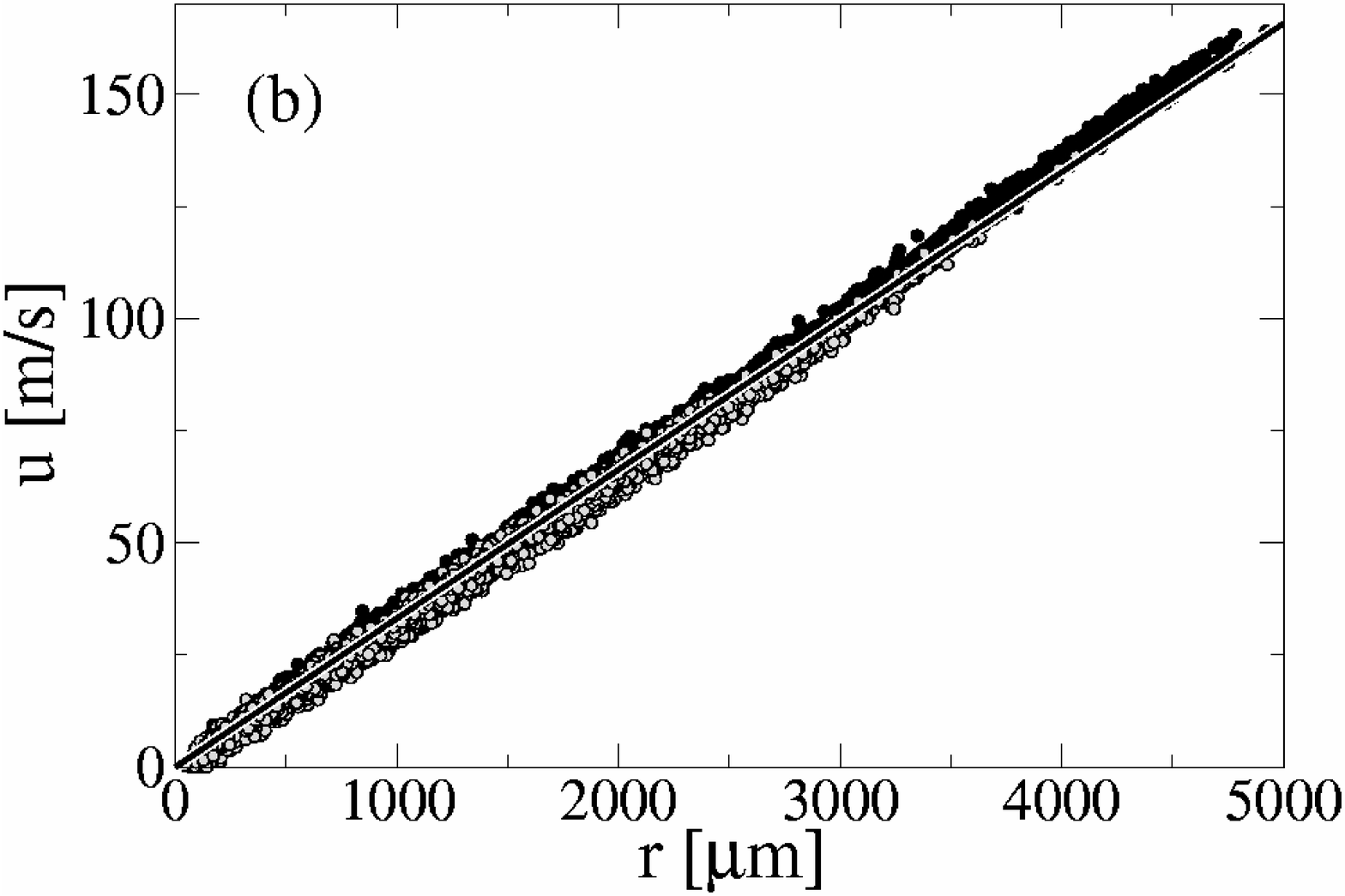,width=8cm}}
\caption{\label{fig5}
Hydrodynamic velocity $u(r)$ of ions (full circles) and atoms (open circles)
at the same times as in Fig.\ \ref{fig4}, compared to the straight-line
assumption of the kinetic model. The dashed line in (a) shows the result of the
particle simulation using a mean-field treatment of the ion-ion interaction.}
\end{figure}
\subsubsection{Evolution of the particle densities}
In the derivation of the kinetic equations (\ref{par}), we have assumed that
the analytical form of the ionic density $\rho_{\rm i}$ remains invariant
during the evolution of
the system and, moreover, that the atoms will have the same distribution.
As the plasma expands, the spatial profile of the ions must deviate from its
original Gaussian shape \cite{Rob03}. This is mainly due to deviations from
quasineutrality, e.g.\ deviations from the linear space dependence of the
outward directed acceleration, at the plasma edge. The influence
of the nonlinear correlation pressure on the density profile
is of minor importance, as can be seen by comparing the solid and dot-dashed
line of Fig.\ \ref{fig4}a in the inner plasma region. As known from earlier
studies of expanding plasmas, based on a mean-field treatment of the particle
interactions \cite{Gur66,Sac85,Rob03}, a sharp spike develops at the plasma
edge, shown by the dot-dashed line in Fig.\ \ref{fig4}a. 
At later times, this spike decays again when the maximum of the hydrodynamic
ion velocity passes the position of the density peak, so that the region of the
peak is depleted. Ultimately, at long times, the plasma approaches a
quasineutral selfsimilar expansion \cite{Sac85}.
From Fig.~\ref{fig4}a it becomes apparent that with IC the peak structure is
less pronounced than in mean-field approximation. This is due
to dissipation caused by ion-ion collisions which are fully taken into account
in the H-MD simulation. As shown in \cite{Sac85}, by adding an ion viscosity
term to the hydrodynamic equations of motion, dissipation tends to 
stabilize the ion density and prevents the occurrence of wavebreaking which was
found to be responsible for the diverging ion
density at the plasma edge in the case of a dissipationless plasma
expansion. Furthermore, the initial correlation heating of the ions largely
increases the thermal ion velocities leading to a broadening of the peak
structure compared to the zero-temperature case.

Apart from the deviations at the plasma edge, the ionic density is
rather well reproduced
by the Gaussian approximation for the spatial distribution. In particular, there
is good agreement between the rms-radii obtained from the MD simulation and the
kinetic model. On the other hand, the spatial distribution of atoms
significantly deviates from that of the ions even
at relatively early times due to the nonlinear density dependence of the
collision rates in Eq.\ (\ref{rate}). However, as also stated in \cite{Rob03},
the total number of atoms is too small to
significantly influence the macroscopic expansion of the system. 

\subsubsection{Spatial dependence of the radial velocities}

\begin{figure}[tb]
\centerline{\psfig{figure=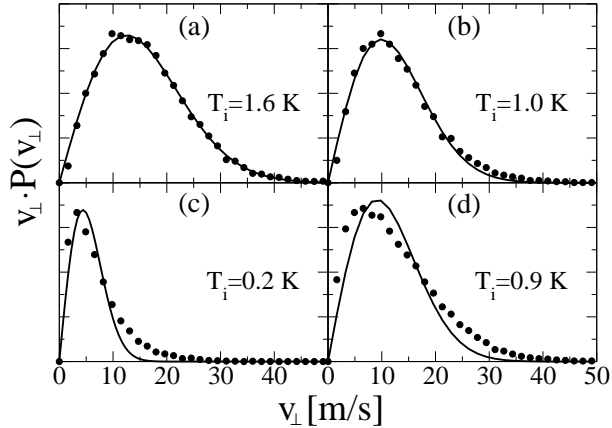,width=8cm}}
\caption{\label{fig6}
Distribution of thermal ionic velocities at $t=0.3\:\mu$s sampled from three
different regions of the plasma: $r\leq 1.3\sigma$ (a), $1.3\sigma<r\leq 2
\sigma$ (b), and $r> 2\sigma$ (c), and from the total plasma volume (d).
The solid lines show a fit to a Maxwell-Boltzmann distribution corresponding to
the temperatures specified in the figure. Initial-state parameters are the same
as in Fig.\ \ref{fig1}.}
\end{figure}

Another assumption used in the derivation of the kinetic model is the
proportionality of the hydrodynamical expansion velocity to the distance
from the center of the plasma cloud, $\mathbf{u} = \gamma \mathbf{r}$, both
for the ions and the atoms. In order to check
this assumption, we have calculated the radial velocity  component
$v_r={\bf{v}}{\bf{r}}/r$ of each particle, which is plotted as
a function of the radial distance from the plasma center in Fig.\ \ref{fig5}.
At an early time the velocity distribution is spread out about its mean value
predicted from the kinetic approach due to the finite ionic temperature.
Note that the expansion is slower near the plasma edge due to the deviation
from quasineutrality as discussed above.
Consequently, the inner part of the plasma
which expands more quickly will catch up with the outer rim, leading to the
formation of the density spike seen in Fig.\ \ref{fig4}a. In the case of the
H-MD simulation the velocity spread, caused by the initial ion heating,
is of the same order of magnitude as the hydrodynamical expansion velocity
itself, leading to a significant broadening of the density spike as discussed
above. At later stages of
the system evolution, the ions cool adiabatically due to the plasma expansion,
and the width of the velocity distribution decreases significantly. Moreover,
as discussed in connection with the decay of the ion density peak in
Fig.\ \ref{fig4}b, the decrease of the ion velocities near the plasma edge
apparent at early times has disappeared.

A comparison with the result of the kinetic model equations (\ref{par})
shows once more that the H-MD simulation not only reproduces the linear
radial dependence of the hydrodynamical velocity, but also yields a
quantitative agreement between both methods.  

\subsubsection{Spatial dependence of the thermal velocities}

\begin{figure}[tb]
\centerline{\psfig{figure=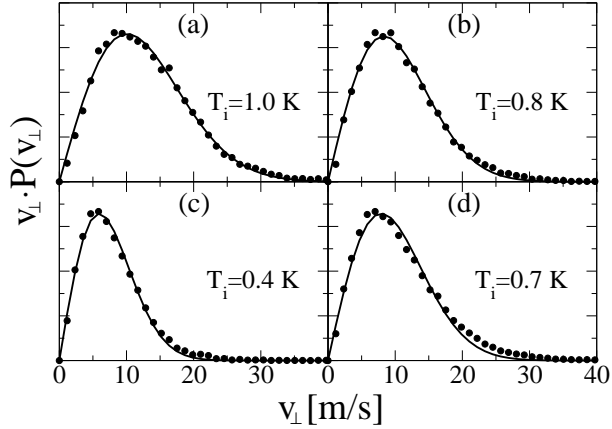,width=8cm}}
\caption{\label{fig7}
Same as Fig.\ \ref{fig6}, but for $t=1.5\:\mu$s.}
\end{figure}
Due to its marginal influence on the plasma expansion dynamics,
the role of the ionic temperature $T_{\rm i}$ for the state of the
system has not been addressed before. In the cold fluid model of
\cite{Rob02,Rob03}, $T_{\rm i}$ has been set to zero in order to follow the
long-time plasma dynamics. However, as stated in
the introduction, one of the motivations of the current type of experiments
was the creation of a strongly coupled plasma. In this context, knowledge
of $T_{\rm i}$ is essential since it directly enters the Coulomb coupling
parameter which determines the state of the plasma. Moreover, the ionic
temperature gives important insight into the relaxation dynamics of the plasma.
For comparing the kinetic model with the H-MD calculations,
the very definition of $T_{\rm i}$
for the MD simulation requires some discussion. As discussed in section II, we
assume a Gaussian velocity distribution, i.e.\ a well-defined temperature
$T_{\rm i}$, for the plasma ions in our kinetic
model. This, of course, is an approximation since the
plasma is not created in an equilibrium state. 
The total kinetic energy of the ions is a sum of the hydrodynamical expansion
energy and a contribution due to the thermal motion of the ions. Since the
hydrodynamical velocity is directed radially (Eq.\ (\ref{sol})),
we determine the thermal energy of the ions from the
average of the velocity component perpendicular to the radial direction
\begin{equation} \label{temp_md}
k_{\rm{B}}T_{\rm{i}}=\frac{m_{\rm{i}}}{2}\left<\left(\frac{{\bf{v}}\times{
\bf{r}}}{r}\right)^2\right>=\frac{m_{\rm{i}}}{2}\left<v_{\perp}^2\right>\;.
\end{equation}
Clearly, such an assignment of a temperature to the average velocity is only
well defined if the ion
velocities $v_{\perp}$ are distributed according to a Maxwell distribution. In
order to check the validity of this requirement, we have sampled the ion
velocity distribution $v_{\perp}$ from three different regions in the plasma:
$r\leq 1.3\sigma$, $1.3\sigma<r\leq 2\sigma$, and $r> 2\sigma$, which have been
chosen so that each region is occupied by approximately the same number of
ions. The resulting distributions are plotted at two different times
$t=0.3\:\mu$s $=1.3 \omega_{\rm{p,i}}^{-1}$ and
$t=1.5\:\mu$s $=7 \omega_{\rm{p,i}}^{-1}$ in Figs.\ \ref{fig6} and
\ref{fig7}, respectively. Additionally, we have fitted a Maxwell-Boltzmann
distribution to the numerical results, formally defining a temperature in the
corresponding plasma region. As can be seen in Fig.\ \ref{fig6}, even at the
very early stage of the plasma evolution the numerical
data is well fitted by an equilibrium distribution in the inner plasma region,
while there are  deviations in the outer region of the plasma since the
relaxation time is longer due to the lower density far away from the plasma
center. However, already after a relatively short time of $t=1.5\:\mu$s the
velocity distributions are well fitted by a Maxwell-Boltzmann distribution in
all three plasma regions (Fig.\ \ref{fig7}). Hence, the ion thermal energy can
be represented by a local temperature $T_{\rm i}(r)$, decreasing with growing
distance from the plasma
center as can be seen from Figs.~\ref{fig6} and \ref{fig7}. This is due to
the fact that the initial heating arises from a
compensation of the negative correlation energy, which is larger in the central
plasma region where the density is higher. However, as becomes apparent by
comparing Figs.\ \ref{fig6} and \ref{fig7}, the thermal energy equilibrates over
the whole plasma volume rather quickly as the system evolves. While the
temperatures defined in the inner and outermost region deviate by a factor of
eight at $t=0.3\:\mu$s, they differ by a factor of two only $1.2\:\mu$s later. 

\begin{figure}[tb]
\centerline{\psfig{figure=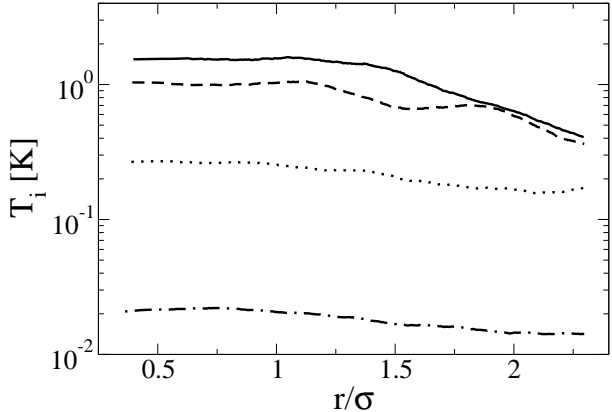,width=8cm}}
\caption{\label{fig8}
Average thermal ionic energy as a function of the distance from the plasma
center at four different times: $t=0.3\:\mu$s (solid), $t=1.5\:\mu$s (dashed),
$t=6.0\:\mu$s (dotted) and $t=25.3\:\mu$s (dot-dashed). Initial-state
parameters are the same as in Fig.\ \ref{fig1}.}
\end{figure} 
Figure \ref{fig8} gives a more detailed account of this equilibration process.
Here, the local ionic temperature is plotted as a
function of the radial distance from the plasma center at four different
times, where $T_{\rm i}(r)$ has been defined from the velocity average of a
shell of 2000 ions with a central shell radius $r$.
The temperature decrease with increasing distance from the center as discussed
above is clearly
visible. Nevertheless, the ion temperature is seen to equilibrate rather
quickly, so that the approximation of a homogeneous ion temperature, used in
the derivation of the kinetic model in section \ref{za}, becomes better and
better at later
times.  Moreover, the numerically calculated distribution of thermal velocities
sampled over the whole plasma volume is well represented by a Maxwell-Boltzmann
distribution with some average temperature intermediate between the
temperatures of the inner and outer region, respectively (Fig.\ \ref{fig7}d).
This shows that the Gaussian phase-space distribution assumed for the ions in
section \ref{za} agrees very well with the results of the MD simulation
averaged over the spatial coordinates, even if the
temperature still shows substantial inhomogeneities. 

\subsection{Spatially averaged ionic observables}

\begin{figure}[tb]
\centerline{\psfig{figure=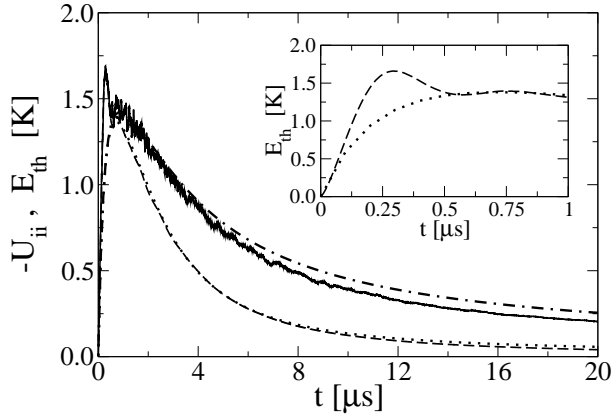,width=8cm}}
\caption{\label{fig9}
Correlation energy (solid line: H-MD simulation, dot-dashed: kinetic
result) and thermal ion kinetic energy (dashed: H-MD, dotted: kinetic).
Initial-state parameters are the same as in Fig.\ \ref{fig1}. The inset
shows the ion thermal energy in the early stage of the relaxation process
with its characteristic transiently oscillatory behavior.}
\end{figure}

As we have demonstrated in section \ref{GlobalResults} the kinetic model
describes the global temporal evolution of the plasma including recombination
quite accurately. From the detailed analysis of the spatially resolved plasma
dynamics in the previous subsection we
may expect that the kinetic model describes spatially averaged
observables, such as the kinetic energy of the expansion, the thermal energy,
and the correlation energy of the plasma quite well. This is indeed the case
over almost the entire evolution time as Fig.\ \ref{fig9} demonstrates 
for the correlation energy and the thermal ion energy.
Only at an early stage of the plasma evolution, differences
between MD simulation and kinetic model are visible, showing that the
correlation-time approximation Eq.\ (\ref{CTA}) does not accurately
describe this early phase of equilibration starting from a completely
uncorrelated state in all details. Since the initial state is very far from
equilibrium, the initial relaxation process is not exponential, as assumed in
the correlation-time approximation Eq.\ (\ref{CTA}). Rather,
it is connected with transient oscillations of the temperature (inset of
Fig.\ \ref{fig9}) which have been found both theoretically
\cite{Zwi,Mor03,PPR04b} and experimentally \cite{Kilpri}.
However,
the timescale of the initial ion heating as well as the maximum temperature
are well reproduced by the simple model. After the system has come
sufficiently close to local equilibrium, the quality of the correlation-time
approximation becomes better and, once
again, close agreement between the two approaches is found, supporting our
argument put forward in the derivation of the kinetic approach in section
II. At later times differences become apparent, which may be attributed to the
fact that the ion relaxation is considerably disturbed by recombination and
ionization events leading to sudden local changes of the charge density,
which is not taken into account by the kinetic model.  

Furthermore, according to both approaches, the correlation energy and the
thermal kinetic energy of the ions
are almost identical roughly to the time where both curves reach their
maximum values, showing that the total correlation energy is completely
converted into thermal kinetic energy of the ions, as expressed by Eq.\
(\ref{parc}). At later times, this additional kinetic energy is transferred to
the outward directed motion of the ions, leading to an indirect enhancement of
the plasma expansion by the development of IC and to adiabatic
cooling of the ions. Therefore, the thermal ion kinetic energy starts to
deviate from the correlation energy as the plasma expansion sets in. 
\section{Conclusions}
In summary, we have presented two different theoretical approaches for the
simulation of ultracold neutral plasmas. First, we have introduced a simple
kinetic model along the lines of \cite{Rob02}, and we have shown how to include
a description
of IC into the model in an approximate way. Moreover, we
have developed a hybrid molecular dynamics approach which allows for an
accurate description of the strongly coupled ion motion on microsecond
timescales, by treating the electronic component as a fluid using an adiabatic
approximation while the ions are fully accounted for on an MD level.

Supporting our results from \cite{PPR03},
both methods show that the inclusion of IC enhances the
number of recombined Rydberg atoms by a few percent, but only slightly
affects the macroscopic expansion dynamics of the plasma itself. As we have
shown, this is due to the fact that mainly the population of very highly
excited states is increased if IC are taken into account,
which have a small binding energy and therefore hardly influence the electron
temperature. 

By comparison of the two methods,
we could show that the simple kinetic description adequately describes the
evolution of global, i.e.\ spatially averaged, plasma observables.
Thus, the kinetic model, which allows for a much faster
computation, may be used to quickly and efficiently scan the vast space of
initial-state parameters, e.g.\ in order to obtain a ``phase diagram'' for
Rydberg gas / plasma systems \cite{FBS}. Moreover, it permits extending the
description of ultracold plasmas to a parameter range where the plasma is so
large that the number of particles ($N_i \agt 10^6$) prohibits an MD
simulation. Maybe even more importantly,
the simple kinetic equations give additional insight into the
dynamics beyond that possible on the basis of MD simulations by providing
simple evolution equations for the macroscopic parameters describing the
plasma state.

On the other hand, spatially resolved quantities such as ionic density, ion
velocities or local temperature show deviations from the behavior
predicted by the simple kinetic model. However,
the developed H-MD approach provides a powerful method for the study of these
quantities, and for the detailed description of the relaxation dynamics of the
strongly coupled ions on a microsecond timescale.
Moreover, it permits the study of scenarios where the ions are so strongly
coupled that Coulomb crystallization occurs \cite{PPR04},
which cannot be described by the kinetic model.

\begin{acknowledgments}
We gratefully acknowledge helpful discussions with T.C.\ Killian and F.\
Robicheaux. This work was supported by the DFG within the Priority
Programme SPP1116 (Grant-No.\ RO1157/4).
\end{acknowledgments}

\begin{appendix}
\section{Derivation of the correlation force}
In this section, the approximation Eq.~(\ref{Fc}) for $\mathbf{F}_{\rm{ii}}$
is derived. We start from Eq.~(\ref{cforce})
\begin{equation} \label{a1b}
\mathbf{F}_{\rm{ii}} = e^2\int \rho_{\rm{i}}({\mathbf{r}}^{\prime}) \,
g_{\rm{ii}}({\mathbf{r}}, {\mathbf{r}}^{\prime}) \, \frac{\mathbf{r}- {
\mathbf{r}}^{\prime}}{\left| {\mathbf{r}}-{\mathbf{r}}^{\prime}\right|^3} \,
d{\mathbf{r}}^{\prime} \;\; ,
\end{equation}
where the explicit expression
$\varphi_{\rm{ii}}=e^2/|{\mathbf{r}}-{\mathbf{r}}^{\prime}|$ for the
inter-ionic Coulomb potential has been inserted.
In general, the correlation function $g_{\rm{ii}}$ is a function of both
coordinates
$\mathbf{r}$ and ${\mathbf{r}}^{\prime}$. However, in the case of a homogeneous
density, $g_{\rm{ii}}$ depends only on the interparticle distance
$x=|{\mathbf{r}}-{\mathbf{r}}^{\prime}|$.
Since the relevant property which distinguishes the two points
${\mathbf{r}}$ and ${\mathbf{r}}^{\prime}$ is the corresponding density (from
the way the plasma is created, no other differences, e.g.\ that part of the
plasma would be in a state with equilibrium correlations while a different
part would be totally uncorrelated, are apparent), it
seems a reasonable approximation to assume that the space dependence of the
correlation function enters only via the densities at
the respective coordinates \cite{Eva79}.
Hence, we write the correlation function as
\begin{equation} \label{a2}
g_{\rm{ii}}({\mathbf{r}},{\mathbf{r}}^{\prime}) \approx g_{\rm{ii}}\left[
\rho_{\rm{i}}({\mathbf{r}}),\rho_{\rm{i}}(
{\mathbf{r}}^{\prime}),|{\mathbf{r}}-{\mathbf{r}}^{\prime}|\right] \;\; .
\end{equation}
With the substitution ${\mathbf{r}}^{\prime}={\mathbf{r}}+{\mathbf{x}}$,
Eq.~(\ref{a1b}) becomes
\begin{equation} \label{a3b}
\mathbf{F}_{\rm{ii}} = -e^2\int \rho_{\rm{i}}({\mathbf{r}}+{\mathbf{x}})
g_{\rm{ii}}\left[\rho_{\rm{i}}({\mathbf{r}}), \rho_{\rm{i}}({\mathbf{r}}+
{\mathbf{x}}),x\right] \frac{\mathbf{x}}{x^3} \, d{\mathbf{x}} \;\; .
\end{equation}
Since the correlation function rapidly decreases for distances $x$ larger than
the correlation length $\lambda_{\rm{c}}$, we may restrict the integration in
Eq.~(\ref{a3b}) to a sphere with a radius of approximately
$\lambda_{\rm{c}}$. If the
plasma density does not vary strongly on the scale of the correlation
length, we may use a linear Taylor expansion of the density
\begin{equation} \label{a4}
\rho_{\rm{i}}\left(\mathbf{r}+\mathbf{x}\right)\approx\rho_{\rm{i}}\left(
\mathbf{r}\right)+
\mathbf{x}\cdot\nabla \rho_{\rm{i}}  \left(\mathbf{r}\right)  
\end{equation}
and the correlation function
\begin{equation} \label{a5}
g_{\rm{ii}}(\rho_{\rm{i}},\rho_{\rm{i}}^{\prime},x)=g_{\rm{ii}}(\rho_{\rm{i}},
\rho_{\rm{i}},x)+ \left. \frac{\partial
g_{\rm{ii}}(\rho_{\rm{i}},\rho_{\rm{i}}^{\prime},x)}{\partial \rho_{
\rm{i}}^{\prime}} \right|_{\rho_{\rm{i}}^\prime=\rho}
\; \left(\mathbf{x}\cdot\nabla\rho_{\rm{i}} \right)  \;\; ,
\end{equation}
where $\rho_{\rm{i}}=\rho_{\rm{i}}(\mathbf{r})$ and $\rho_{\rm{i}}^{\prime}=
\rho_{\rm{i}}(\mathbf{r}+\mathbf{x})$.

Substitution of Eq.\ (\ref{a4}) and Eq.~(\ref{a5})
into Eq.~(\ref{a3b}) and keeping only terms up to linear order in
${\bf{x}}\cdot\nabla\rho_{\rm{i}}$ yields 
\begin{widetext}
\begin{equation} \label{a6}
\mathbf{F}_{\rm{ii}}=-e^2\left(
\int \frac{\mathbf{x}}{x^3} \; \rho_{\rm{i}} \; g_{\rm{ii}}(\rho_{\rm{i}},x)\;
d\mathbf{x} + \int \frac{\mathbf{x}}{x^3} \; g_{\rm{ii}}(\rho_{\rm{i}},x)\;
\left(\mathbf{x}\cdot \nabla\rho_{\rm{i}}\right) \; d\mathbf{x} \right. + 
\left. \frac{1}{2} \int \frac{\mathbf{x}}{x^3} \; \rho_{\rm{i}} \;
\frac{\partial g_{\rm{ii}}(\rho_{\rm{i}},x)}{\partial \rho_{\rm{i}}}\;\left(
\mathbf{x}\cdot \nabla\rho_{\rm{i}}\right)\; d\mathbf{x} \right) \; ,
\end{equation}
where $g_{\rm{ii}}(\rho_{\rm{i}},x) \equiv \left. g(\rho_{\rm{i}},
\rho_{\rm{i}}^\prime, x) \right|_{\rho_{\rm{i}}^\prime =
\rho_{\rm{i}}}$ and we have used the relation
\begin{equation} \label{ag}
\frac{\partial}{\partial \rho_{\rm{i}}} \left( \left. g_{\rm{ii}}(\rho_{\rm{i}},
\rho_{\rm{i}}^\prime, x)
\right|_{\rho_{\rm{i}}^\prime = \rho_{\rm{i}}} \right) = \left. \frac{\partial
g(\rho_{\rm{i}},
\rho_{\rm{i}}^\prime, x)}{\partial \rho_{\rm{i}}} \right|_{\rho_{\rm{i}}^\prime
= \rho_{\rm{i}}} +
\left. \frac{\partial g_{\rm{ii}}(\rho_{\rm{i}}, \rho_{\rm{i}}^\prime,
x)}{\partial \rho_{\rm{i}}^\prime}
\right|_{\rho_{\rm{i}}^\prime = \rho_{\rm{i}}} = 2 \left. \frac{\partial
g(\rho_{\rm{i}}, \rho_{\rm{i}}^\prime,
x)}{\partial \rho_{\rm{i}}^\prime} \right|_{\rho_{\rm{i}}^\prime =
\rho_{\rm{i}}} \;\; ,
\end{equation}
\end{widetext}
which follows from the symmetry of $g_{\rm{ii}}$ under particle exchange,
i.e., under exchange
of $\rho_{\rm{i}}$ and $\rho_{\rm{i}}^{\prime}$.
Since the integrand of the first integral in Eq.~(\ref{a6}) is an odd function
in ${\bf{x}}$ the first term vanishes. The second term yields after some
manipulations
\begin{equation}
e^2 \int \frac{\mathbf{x}}{x^3} \; g_{\rm{ii}}(\rho_{\rm{i}},x) \;\left(
\mathbf{x} \cdot \nabla \rho_{\rm{i}}\right)\; d\mathbf{x} 
=\frac{e^2}{3}\; \nabla\rho_{\rm{i}}\; \int\frac{g_{\rm{ii}}(\rho_{\rm{i}},
x)}{x} \; d\mathbf{x}\;. 
\end{equation}
Analogously, the third term can be written as
\begin{equation}
\frac{e^2}{2} \int \frac{\mathbf{x}}{x^3} \; \rho_{\rm{i}} \; \frac{\partial
g_{\rm{ii}}(\rho_{\rm{i}},x)}{\partial \rho_{\rm{i}}}\;\left(\mathbf{x} \cdot
\nabla\rho_{\rm{i}}\right) \; d\mathbf{x}=\frac{e^2}{6} \rho_{\rm{i}}\; \nabla\rho_{\rm{i}}\frac{\partial}{\partial
\rho_{\rm{i}}} \int\frac{g_{\rm{ii}}(\rho_{\rm{i}},x)}{x} \; d\mathbf{x}\;,
\end{equation}
which together leads to
\begin{eqnarray} \label{a7}
\mathbf{F}_{\rm{ii}}=-\frac{e^2}{6}\;\nabla\rho_{\rm{i}}\;\left[\int\frac{
g_{\rm{ii}}(\rho_{\rm{i}},x)}{x} \; d\mathbf{x}+\; \frac{\partial}{\partial \rho_{\rm{i}}}\left(\rho_{\rm{i}}\int
\frac{g_{\rm{ii}}(\rho_{\rm{i}},x)}{x}\;d\mathbf{x} \right) \right].
\end{eqnarray}
Finally, substitution of the definition of the correlation energy $u$ as given
by Eq.~(\ref{uc}) yields the result Eq.~(\ref{Fc}).
An analogous calculation for the expression of the correlation energy
Eq.~(\ref{Uc2}) leads to the familiar LDA result \cite{Eva79} 
\begin{eqnarray} \label{a8}
U_{\rm{ii}}&=&\frac{e^2}{2N_{\rm{i}}}\int \rho_{\rm{i}}({\mathbf{r}})\rho_{
\rm{i}}({\mathbf{r}}^{\prime}) \, g_{\rm{ii}}({\mathbf{r}},
{\mathbf{r}}^{\prime}) \, \frac{1}{\left| {\mathbf{r}}-{\mathbf{r}}^{\prime}
\right|} \, d{\mathbf{r}}^{\prime}d{\mathbf{r}}\nonumber \\
&=&N_{\rm{i}}^{-1}
\int\rho_{\rm{i}}u_{\rm{ii}}\; d\mathbf{r}\;.
\end{eqnarray}
\end{appendix}

\end{document}